\documentclass[pre, twocolumn, showkeys, amsmath, amssymb]{revtex4}

\usepackage{graphicx}

\begin{document}

\title{A comparison of sexual and asexual replication strategies in a simplified model based on the yeast life cycle}

\author{Emmanuel Tannenbaum}
\email{emanuelt@bgu.ac.il}
\affiliation{Department of Chemistry, Ben-Gurion University of the Negev, Be'er-Sheva, Israel}

\begin{abstract}

This paper develops simplified mathematical models describing the mutation-selection balance for the asexual and sexual replication pathways in {\it Saccharomyces cerevisiae}, or Baker's yeast.  The simplified models are based on the single-fitness-peak approximation in quasispecies theory.  We assume diploid genomes consisting of two chromosomes, and we assume that each chromosome is functional if and only if its base sequence is identical to some master sequence.  The growth and replication of the yeast cells is modeled as a first-order process, with first-order growth rate constants that are determined by whether a given genome consists of zero, one, or two functional chromosomes.  In the asexual pathway, we assume that a given diploid cell divides into two diploids.  For the sake of generality, our model allows for recombination.  In the sexual pathway, we assume that a given diploid cell divides into two diploids, each of which then divide into two haploids.  The resulting four haploids enter a haploid pool, where they grow and replicate until they meet another haploid with which to fuse.  In the sexual pathway, we consider two mating strategies:  (1)  A selective strategy, where only haploids with functional chromosomes can fuse with one another;  (2)  A random strategy, where haploids randomly fuse with one another.  When the cost for sex is low, we find that the selective mating strategy leads to the highest mean fitness of the population, when compared to all of the other strategies.  We also show that, at low to intermediate replication fidelities, sexual replication with random mating has a higher mean fitness than asexual replication, as long as the cost for sex is low.  If the fitness penalty for having a defective chromosome is sufficiently high and the cost for sex sufficiently low, then at low to intermediate mutation rates the random mating strategy has a mean fitness that is a factor of $ \sqrt{2} $ larger than the asexual mean fitness.  This is consistent with previous work suggesting that sexual replication is advantageous at high population densities, low replication rates, and intermediate replication fidelities.  The results of this paper also suggest that {\it S. cerevisiae} switches from asexual to sexual replication when stressed, because stressful growth conditions provide an opportunity for the yeast to clear out deleterious mutations from their genomes.  That being said, our model does not contradict theories for the evolution of sex that argue that sex evolved because it allows a population to more easily adapt to changing conditions.

\end{abstract}


\maketitle

\section{Introduction}

The evolution and maintenance of sexual replication is one of the central problems in evolutionary biology \cite{Bell:82, Williams:75, Smith:78, Michod:95, Hurst:96}.  While sex is observed to be the preferred, and often the only, replication strategy in the more complex, multicellular organisms, various forms of sexual replication are known to occur in single-celled organisms as well.  

The ubiquity of sexual replication suggests that the replication strategy provides a selective advantage over asexual replication.  The two basic theories for the selective advantage for sex are that sex provides a mechanism for removing deleterious genes from a population \cite{Michod:95, Bernstein:85, Muller:64, Otto:06, Otto:05}, and that sex allows for faster adaptation in dynamic environments \cite{Bell:82, Hamilton:90, Otto:06, Otto:05}.  While these two basic theories are often presented as mutually contradictory, this is not necessarily the case:  That is, sex may provide a mechanism for removing deleterious genes from a population, and this same mechanism may also be responsible for allowing faster adaptation in dynamic environments.

The various theories for the existence of sex are incomplete, in that they provide an explanation for the selective advantage of the strategy, but do not explain why some organisms are essentially asexual, with some ability for recombination, while other organisms replicate exclusively sexually.

In a recent set of papers \cite{Tannenbaum:06, TannFon:06, LeeTann:07, Tannenbaum:07}, Tannenbaum, Fontanari, and Lee argued that sex is generally favored in more slowly replicating organisms, since in this case the time (and energy) costs associated with finding a recombination partner are small compared to the characteristic growth time of the organism.  For more quickly replicating organisms, the time costs associated with sex become sufficiently large so that the asexual strategy becomes advantageous.  

The results of the models presented by Tannenbaum, Fontanari, and Lee are broadly consistent with observation.  However, these models have the drawback that they assume highly simplified sexual replication pathways that do not exactly correspond to the sexual replication pathways in actual organisms.  These pathways were considered because they were analytically solvable, and yet deemed a sufficiently good approximation of actual pathways to yield biologically relevant results.  Nevertheless, a proper understanding of the evolutionary basis for sex will only be achieved when the asexual and sexual replication pathways of actual organisms are analyzed.

In this paper, we seek to develop and analyze more realistic models comparing asexual and sexual replication strategies in unicellular organisms.  In this vein, we are interested in constructing models based on the asexual and sexual replication pathways in {\it Saccharomyces cerevisiae}, or Baker's yeast.  

{\it S. cerevisiae} is a model organism that is used to investigate numerous fundamental problems in biology.  Among these is the evolutionary basis for sexual replication.  The reason for this is that {\it S. cerevisiae}, when stressed, engages in a form of sexual replication that is essentially the unicellular analogue of the gamete-based sexual replication pathway in more complex organisms.  Because {\it S. cerevisiae} is capable of both asexual and sexual replication, understanding the reason for this organism to choose one replication strategy over another in a given situation will provide a clearer picture of the advantages and disadvantages of sexual replication over asexual replication.

This paper is organized as follows:  In the following section (Section II), we develop and analyze a model describing the evolutionary dynamics of a population of {\it S. cerevisiae} replicating asexually.  In Section III, we develop and analyze two models describing the evolutionary dynamics of a population of {\it S. cerevisiae} replicating sexually.  The first model assumes a selective mating strategy, whereby only viable haploids may recombine with one another, while the second model assumes a random mating strategy.  We find that when the cost for sex, as measured by the characteristic haploid fusion time, is negligible, then the selective mating strategy is advantageous over asexual replication for all replication fidelities, while the random mating strategy is only guaranteed to be advantageous over asexual replication at low to intermediate replication fidelities.  In Section IV, we consider the case of self-fertilization, and show that, while such a strategy can be advantageous over pure asexual replication, the sexual replication strategies of Section III are still advantageous at low to intermediate replication fidelities, as long as the cost for sex is negligible.  Finally, in Section V, we summarize the main results of this paper and describe plans for future research.  In particular, we discuss how the results in this paper provide a plausible explanation for the benefits of sexual replication as a stress response in {\it S. cerevisiae}.  

It should be noted that the models we consider in this paper are highly simplified, in that we assume genomes consisting of only two chromosomes, and a fitness landscape that is analogous to the single-fitness-peak landscape for single-stranded genomes.  Therefore, while these models are not intended to be quantitative at this stage, they are the two-chromosomed, single-fitness-peak analogues of the actual replication dynamics of {\it S. cerevisiae}, and are therefore more realistic than previous models of unicellular sexual replication.

\section{Asexual Replication}

\subsection{Definitions}

We begin our analysis by considering the evolutionary dynamics of a unicellular population replicating asexually.  For simplicity, we assume that the genome of each cell consists of two chromosomes, and that, by analogy with the single-fitness-peak landscape in single-chromosomed models, a given chromosome is functional if and only if it is equal to some master sequence, denoted $ \sigma_0 $.  It follows that a given genome consists of either zero, one, or two functional chromosomes.  

We assume that replication is characterized by first-order kinetics, and that the first-order growth rate constant depends on the number of functional chromosomes in the genome.  We let $ \kappa_{vv} $, 
$ \kappa_{vu} $, and $ \kappa_{uu} $ denote the first-order growth rate constants of cells with genomes consisting of two, one, and zero functional chromosomes, respectively.  Here, ``v" stands for ``viable", and ``u" stands for ``unviable".  Naturally, we assume that $ \kappa_{vv} \geq \kappa_{vu} \geq \kappa_{uu} $.  Furthermore, for the remainder of this paper, we will assume that $ \kappa_{uu} = 0 $, and we will also define $ \alpha = \kappa_{vu}/\kappa_{vv} $, so that $ \alpha $ is the fitness penalty associated with having a defective chromosome.

We also let $ n_{vv} $, $ n_{vu} $, and $ n_{uu} $ denote the number of organisms with two, one, and zero functional chromosomes respectively.  We may then define the population fractions $ x_{vv} $, $ x_{vu} $, and $ x_{uu} $ as $ x_{vv} = n_{vv}/n $, $ x_{vu} = n_{vu}/n $, $ x_{uu} = n_{uu}/n $, where $ n \equiv n_{vv} + n_{vu} + n_{uu} $ is the total population.

When a cell divides into two new diploids, the two parent chromosomes first replicate, and then distribute themselves evenly between the two cells.  We assume that replication is not error-free, but that a given chromosome has a probability $ p $ of being replicated correctly.  If we neglect backmutations, then a non-functional chromosome cannot produce a functional daughter.  Therefore, a ``v" chromosome produces a ``v" daughter with probability $ p $, a ``u" daughter with probability $ 1 - p $, while a ``u" chromosome produces a ``u" daughter with probability $ 1 $.

Finally, to allow for mitotic recombination within the limits of this model, we do not assume that the two chromosomes of a given parent-daughter pair segregate into distinct cells.  Rather, if we tag one of the parent chromosomes with a ``1", the other parent with a ``2", the daughter of the ``1" chromosome with a ``3", and the daughter of the ``2" chromosome with a ``4", then the ``1" chromosome has a certain probability of co-segregating with any of the other chromosomes.  Thus, we let $ r_{12} $ denote the probability that the two parent chromosomes co-segregate into one of the daughter cells, so that the two daughter chromosomes co-segregate into the other daughter cell; we let $ r_{13} $ denote the probability that parent chromosome ``1" co-segregates with its daughter, so that parent chromosome ``2" co-segregates with its daughter; and we let $ r_{14} $ denote the probability that parent chromosome ``1" co-segregates with the daughter of parent chromosome ``2".  Note that $ r_{12} + r_{13} + r_{14} = 1 $.

\subsection{Population genetics equations}

With the recombination probabilities $ r_{12} $, $ r_{13} $, $ r_{14} $ and replication fidelity $ p $ in hand, we may compute the probabilities of the various replication pathways, as shown in Figure 1.  This leads to the following system of differential equations governing the time evolution of the various population fractions:

\begin{eqnarray}
&  &
\frac{d x_{vv}}{dt} = [\kappa_{vv} (2 p + r_{12} (1 - p)^2 - 1) - \bar{\kappa}(t)] x_{vv} 
\nonumber \\
&   &
+ \kappa_{vu} r_{13} p x_{vu}
\nonumber \\
&  &
\frac{d x_{vu}}{dt} = [\kappa_{vu} p (1 - 2 r_{13}) - \bar{\kappa}(t = \infty)] x_{vu} 
\nonumber \\
&  &
+ 2 \kappa_{vv} x_{vv} (1 - p) [p + (1 - r_{12}) (1 - p)]
\nonumber \\
&  &
\frac{d x_{uu}}{dt} =  - \bar{\kappa}(t) x_{uu} + \kappa_{vu} x_{vu} [1 - (1 - r_{13}) p] 
\nonumber \\
&  &
+ \kappa_{vv} x_{vv} r_{12} (1 - p)^2
\end{eqnarray}

We have introduced an additional parameter, $ \bar{\kappa}(t) $, which is the mean fitness of the population and is defined as $ \bar{\kappa}(t) = (1/n) (dn/dt) = \kappa_{vv} x_{vv} + \kappa_{vu} x_{vu} $.  The mean fitness is the per capita growth rate of the population, and therefore measures the first-order growth rate constant of the population as a whole.

\begin{figure}
\includegraphics[width = 0.9\linewidth, angle = 0]{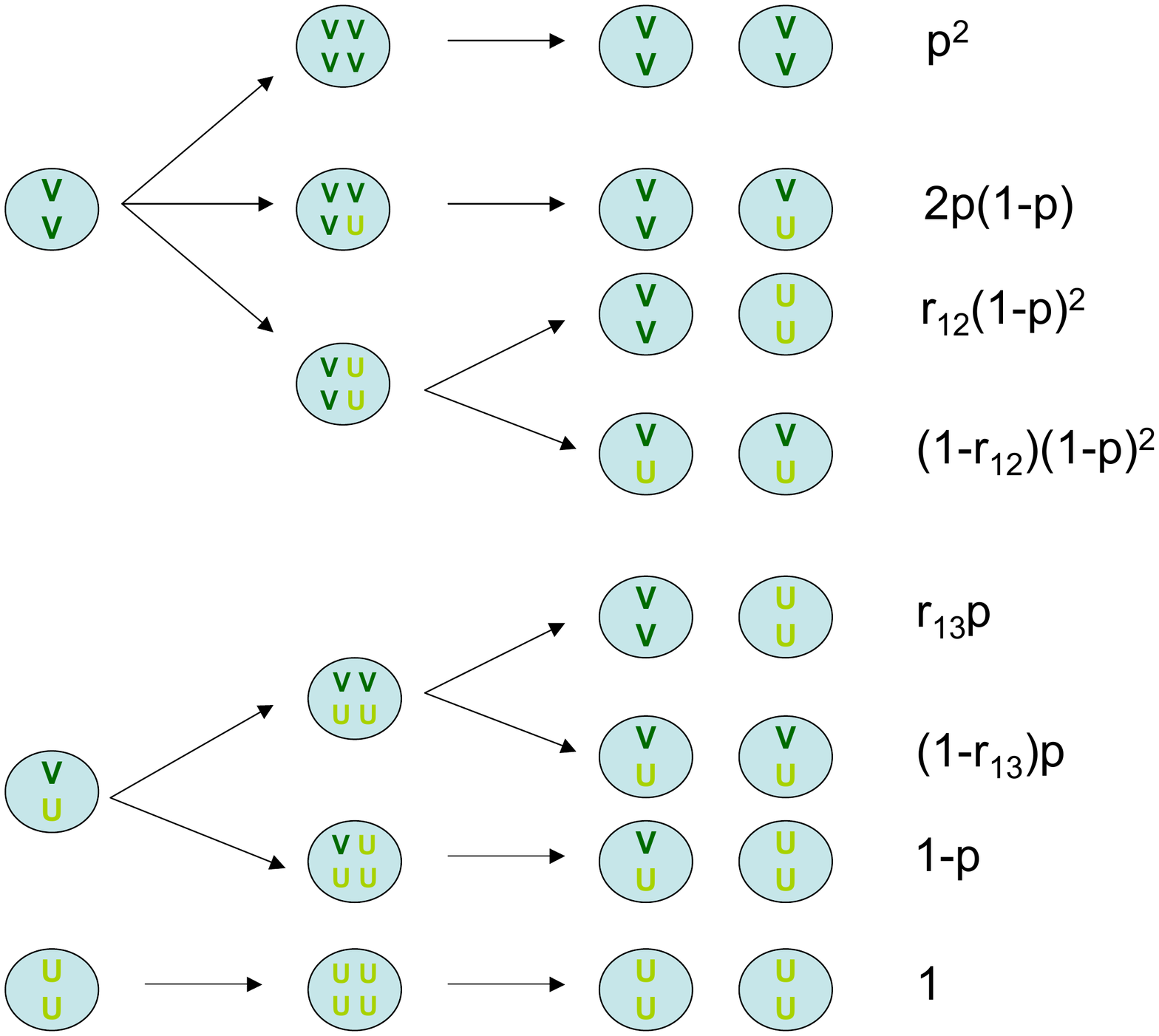}
\caption{(Color online) Illustration of the various replication pathways and their associated probabilities, as a function of the recombination probabilities $ r_{12} $, $ r_{13} $, and $ r_{14} $.}
\end{figure}

In the {\it group selection} approach that will be adopted in this paper, the central object of study is the steady-state mean fitness of a population.  The reason for this is that if two or more populations are growing exponentially, the population will the largest mean fitness will drive the others to extinction.  Therefore, for a given set of parameters, the replication strategy with the largest steady-state mean fitness is the advantageous one.

\subsection{Mutation-selection balance and the steady-state mean fitness}

It is a standard result from population genetics that the above system of differential equations will converge to a steady-state, corresponding to what is known as a {\it mutation-selection balance}.  Therefore, to determine the mutation-selection balance, we set the left-hand-side of the above system of equations to $ 0 $, giving that,
\begin{eqnarray}
&  &
0 = [\kappa_{vv} (2 p + r_{12} (1 - p)^2 - 1) - \bar{\kappa}(t = \infty)] x_{vv} + \kappa_{vu} r_{13} p x_{vu}
\nonumber \\
&  &
0 = [\kappa_{vu} p (1 - 2 r_{13}) - \bar{\kappa}(t = \infty)] x_{vu} 
\nonumber \\
&  & 
+ 2 \kappa_{vv} x_{vv} (1 - p) [p + (1 - r_{12}) (1 - p)]
\nonumber \\
&  &
0 = -\bar{\kappa}(t = \infty) x_{uu} + \kappa_{vu} x_{vu} [1 - (1 - r_{13}) p]
\nonumber \\
&  &
+ \kappa_{vv} x_{vv} r_{12} (1 - p)^2
\end{eqnarray}

Solving the first equation for $ x_{vv} $ in terms of $ x_{vu} $ gives,
\begin{equation}
x_{vv} = \frac{\kappa_{vu} r_{13} p x_{vu}}{\bar{\kappa}(t = \infty) - \kappa_{vv} (2 p + r_{12} (1 - p)^2 - 1)}
\end{equation}

Substituting into the second equation gives,
\begin{eqnarray}
0 
& = & 
x_{vu} 
[\kappa_{vu} p (1 - 2 r_{13}) - \bar{\kappa}(t = \infty) 
\nonumber \\
&   &
+ 2 (1 - p) [p + (1 - r_{12}) (1 - p)]
\times \nonumber \\
&   &
\frac{\kappa_{vv} \kappa_{vu} r_{13} p}{\bar{\kappa}(t = \infty) - \kappa_{vv} (2 p + r_{12} (1 - p)^2 - 1)}
\end{eqnarray}
and so, we either have that $ x_{vu} = x_{vv} = 0 \Rightarrow x_{uu} = 1 $, so that $ \bar{\kappa}(t = \infty) = 0 $, or $ \bar{\kappa}(t = \infty) $ is the solution to the quadratic,
\begin{eqnarray}
0 
& = & 
\bar{\kappa}(t = \infty)^2 - [\kappa_{vv} (2 p + r_{12} (1 - p)^2 - 1) 
\nonumber \\
&   &
+ \kappa_{vu} p (1 - 2 r_{13})] \bar{\kappa}(t = \infty) 
\nonumber \\
&   &
+ \kappa_{vv} \kappa_{vu} p [2 (1 - r_{13}) p + r_{12} (1 - p)^2 - 1]
\end{eqnarray}

Dividing both sides by $ \kappa_{vv}^2 $, we obtain that the normalized steady-state mean fitness, $ \phi \equiv \bar{\kappa}(t = \infty)/\kappa_{vv} $, satisfies the quadratic,
\begin{eqnarray}
0 
& = &
\phi^2 - [2 p + r_{12} (1 - p)^2 - 1 + \alpha p (1 - 2 r_{13})] \phi 
\nonumber \\
&   &
+ \alpha p [2 (1 - r_{13}) p + r_{12} (1 - p)^2 - 1]
\end{eqnarray}
as long as $ x_{vv} $ and $ x_{vu} $ are not both $ 0 $, which is equivalent to the statement that $ \phi > 0 $.

The quadratic in Eq. (6) has two solutions.  To determine which of these solutions corresponds to the mean fitness of the population, we examine the two solutions for both $ p = 0 $ and $ p = 1 $.  When $ p = 0 $, Eq. (6) becomes,
\begin{equation}
0 = \phi [\phi + (1 - r_{12})]
\end{equation}
which admits the solutions $ \phi = 0 $ and $ \phi = r_{12} - 1 \leq 0 $.  The physical solution is $ \phi = 0 $, since a negative mean fitness is impossible for this model (this model does not include death).  

When $ p = 1 $, Eq. (6) becomes,
\begin{equation}
0 = [\phi - \alpha (1 - 2 r_{13})][\phi - 1]
\end{equation}
which admits the solutions $ \phi = \alpha (1 - 2 r_{13}) \leq 1 $, and $ \phi = 1 $.  The physical solution is $ \phi = 1 $, since a population should consist entirely of the wild-type when replication is error-free.

Note that for both $ p = 0 $ and $ p = 1 $, the value of $ \phi $ is the larger of the two roots of the quadratic polynomial in Eq. (6).  By continuity, we expect that the value of $ \phi $ for all $ p \in [0, 1] $ is given by the larger of the two roots of the quadratic polynomial in Eq. (6).  

We may also evaluate $ \phi $ for $ \alpha = 0 $ and $ \alpha = 1 $.  When $ \alpha = 0 $, we obtain that $ \phi = \max\{ 2 p + r_{12} (1 - p)^2 - 1, 0 \} $, while when $ \alpha = 1 $ we have that $ \phi $ is the solution to,
\begin{equation}
0 = [\phi - p] [\phi - (2 (1 - r_{13}) p + r_{12} (1 - p)^2 - 1)]
\end{equation}
which admits the solutions $ \phi = p $ and $ \phi = 2 (1 - r_{13}) p + r_{12} (1 - p)^2 - 1 $.  Because
$ 2 (1 - r_{13}) p + r_{12} (1 - p)^2 - 1 \leq 2 p + (1 - p)^2 - 1 = p^2 \leq p $, we have that $ \phi = p $ for $ \alpha = 1 $.

When $ r_{13} = 0 $, we obtain that $ \phi $ is the solution to,
\begin{equation}
0 = [\phi - \alpha p] [\phi - (2 p + r_{12} (1 - p)^2 - 1]
\end{equation}
so that $ \phi = \max \{2 p + r_{12} (1 - p)^2 - 1, \alpha p\} $.

Finally, when $ \alpha, p, r_{13} \in (0, 1) $, we can show that $ \phi \in (\alpha p, p) $.  To do this, note first that the polynomial of Eq. (6) goes to $ \infty $ as $ \phi \rightarrow -\infty $.  When $ \phi = \alpha p $, the polynomial evaluates to $ - 2 r_{13} \alpha (1 - \alpha) p^2 < 0 $, and so, by the Intermediate Value Theorem, the polynomial of Eq. (6) has a root in $ (-\infty, \alpha p) $.  

When $ \phi = p $, the polynomial evaluates to $ p (1 - p) [1 + \alpha - (1 - \alpha) r_{12} (1 - p)] > 0 $,
and so, again by the Intermediate Value Theorem, the polynomial has a root in $ (\alpha p, p) $.  Because the polynomial of Eq. (6) has only two roots, and because the larger of the two roots is the value of $ \phi $, we have that $ \phi \in (\alpha p, p) $ whenever $ \alpha, p, r_{13} \in (0, 1) $.

\subsection{Limiting behaviors of $ \phi $}

We may determine $ (d \phi/d p)_{p = 0, 1} $ by differentiating both sides of Eq. (6) and substituting $ p = 0 $ and $ p = 1 $, respectively.  For $ p = 0 $, we obtain, after some manipulation,
\begin{equation}
(\frac{d \phi}{d p})_{p = 0} = \alpha
\end{equation}
while when $ p = 1 $ we obtain,
\begin{equation}
(\frac{d \phi}{d p})_{p = 1} = 2 \frac{1 - \alpha (1 - r_{13})}{1 - \alpha (1 - 2 r_{13})}
\end{equation}

Note then that when $ p $ is close to $ 0 $, we have that $ \phi \approx \alpha p $, independently of the values of $ r_{12} $, $ r_{13} $, and $ r_{14} $.

\section{Sexual Replication}

\subsection{Definitions}

The sexual pathway in yeast involves the division of a diploid into two daughter diploid cells, each of which divides again to produce a total of four haploids.  The four haploids then enter a haploid pool, where they continue replicating until they encounter another haploid with which they fuse to form a diploid that may grow and then repeat the cycle.  This process is illustrated in Figure 2.

\begin{figure}
\includegraphics[width = 0.9\linewidth, angle = 0]{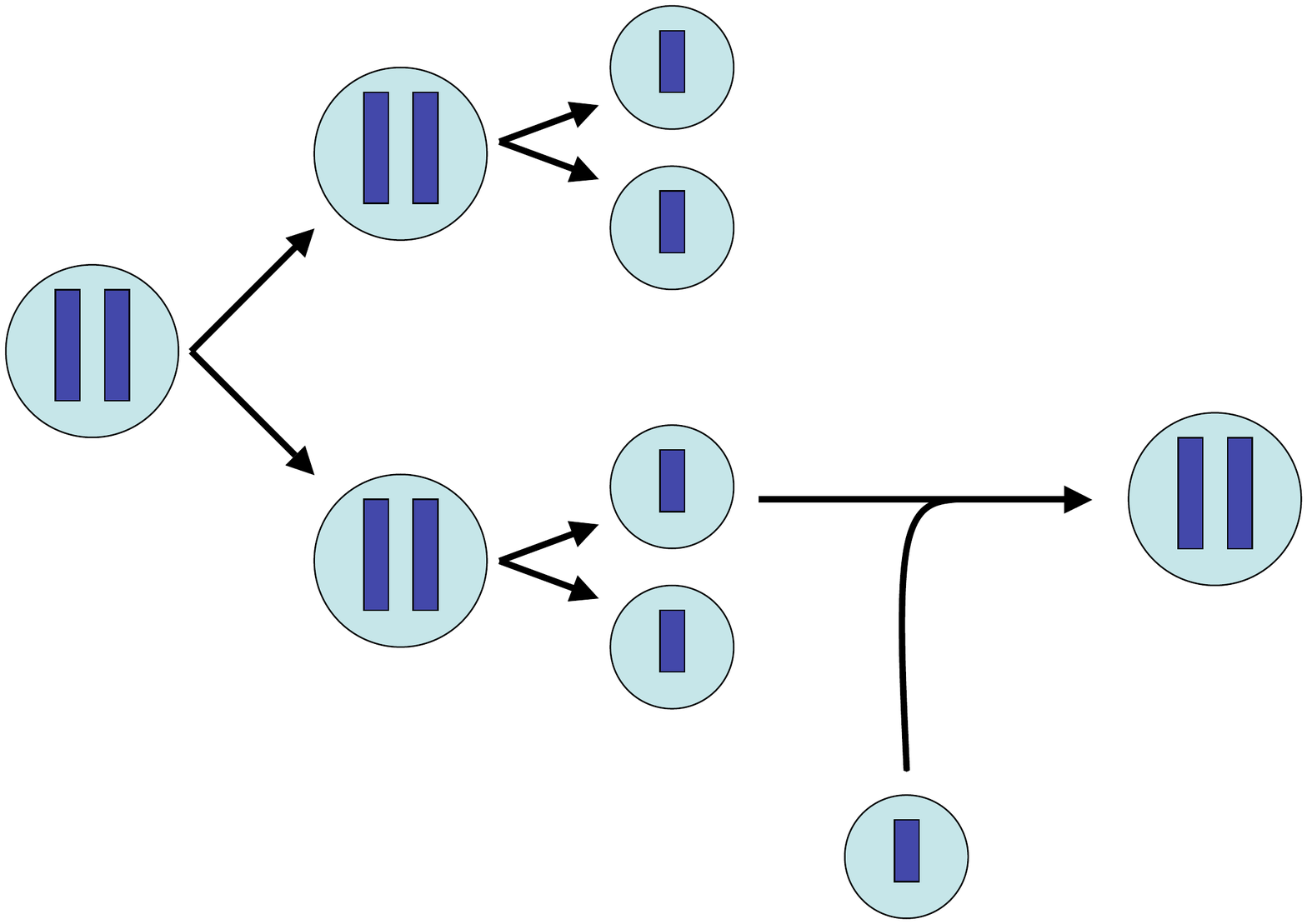}
\caption{Illustration of the sexual replication pathway in {\it S. cerevisiae}.}
\end{figure}

The haploid cells of {\it S. cerevisiae} belong to one of two distinct mating types, termed $ a $ and $ \alpha $.  A given haploid can only fuse with another haploid of a different mating type, so that only $ a-\alpha $ matings are possible.  However, when a haploid of a given mating type divides, the daughter can sometimes spontaneously switch mating types.  

To develop a model for sexual replication in yeast, we need to define a few additional quantities.  To this end, we let $ \kappa_v $ and $ \kappa_u $ denote the first-order growth rate constants of viable and unviable haploid yeast, respectively.  As with diploid yeast, we assume that $ \kappa_u = 0 $.  We let $ n_{v, a} $, $ n_{v, \alpha} $, $ n_{u, a} $, $ n_{u, \alpha} $ denote the number of viable $ a $, viable $ \alpha $, unviable $ a $, and unviable $ \alpha $ haploids in the population.  We also define $ n = n_{vv} + n_{vu} + n_{uu} + (1/2) [n_{v, a} + n_{v, \alpha} + n_{u, a} + n_{u, \alpha}] $.  Finally, we define the population fractions $ x_{vv} = n_{vv}/n $, $ x_{vu} = n_{vu}/n $, $ x_{uu} = n_{uu}/n $, $ x_{v, a} = (1/2) n_{v, a}/n $, $ x_{v, \alpha} = (1/2) n_{v, \alpha}/n $, $ x_{u, a} = (1/2) n_{u, a}/n $, $ x_{u, \alpha} = (1/2) n_{u, \alpha}/n $.

We assume that haploid fusion may be modeled as a second-order rate process with rate constant 
$ \gamma $.  We also assume that, as the population grows, the system volume expands to maintain a constant population density $ \rho $, defined as $ \rho = n/V $.  Finally, we let $ p _m $ denote the probability that when a haploid cell replicates, the daughter will be of the opposite mating type from the parent (because {\it S. cerevisiae} replicates by budding, it is possible to canonically define a single daughter cell).

We consider two distinct mating strategies:  (1)  Selective mating, where we assume that only viable haploids may fuse with one another.  (2)  Random mating, where all haploids may participate in the mating process, and where haploids do not exhibit any kind of preference for a certain genome type.

The reasoning behind considering a selective mating strategy is two-fold:  On the one hand, it may be argued that haploids with unviable genomes are simply physically unable to participate in the mating process.  On the other hand, it may be that the viable haploids have some mechanism for discerning the genome of another haploid (say by reading what are known as {\it indicator traits}) \cite{Andersson:94}, and essentially choose to only fuse with another viable haploid.

The reasoning behind considering a random mating strategy is that in practice, it is very difficult to determine another organism's genome.  Indicator traits (i.e. the phenotype) are not in one-to-one correspondence with the genome (i.e. the genotype).  Furthermore, the actual process of determining an organism's fitness takes time and consumes energy, all of which incur additional fitness penalties on the sexual replication strategy.  These two factors can make the selective mating strategy highly impractical if not impossible.  The random mating strategy does not suffer from the difficulties of the selective mating strategy, and so it is important to consider its selective advantage, if any, over asexual replication, if we are to draw biologically relevant conclusions from our two-chromosome model.

The difficulty and the fitness penalty associated with reading an organismal genome exactly on the one hand, and the benefit of fusing with a fit haploid, likely means that actual mating strategies in real organisms lie somewhere between the two extremes being considered here.  Nevertheless, with the highly simplified models we are considering, the selective and the random mating strategies are the most natural ones to initially study.

\subsection{Selective mating}

The population genetics equations for the selective mating strategy may be readily shown to be,
\begin{eqnarray}
&   &
\frac{d x_{vv}}{dt} = -[\kappa_{vv} + \bar{\kappa}(t)] x_{vv} + \gamma \rho x_v^2
\nonumber \\
&   &
\frac{d x_{vu}}{dt} = -[\kappa_{vu} + \bar{\kappa}(t)] x_{vu}
\nonumber \\
&   &
\frac{d x_{uu}}{dt} = -\bar{\kappa}(t) x_{uu}
\nonumber \\
&   &
\frac{d x_v}{dt} = -\bar{\kappa}(t) x_v + \kappa_{vv} x_{vv} (1 + p) + \frac{1}{2} \kappa_{vu} x_{vu} (1 + p)
\nonumber \\
&   &
+ \kappa_v x_v p - \gamma \rho x_v^2
\nonumber \\
&   &
\frac{d x_u}{dt} = -\bar{\kappa}(t) x_u + \kappa_{vv} n_{vv} (1 - p) + \frac{1}{2} \kappa_{vu} n_{vu} (3 - p) 
\nonumber \\
&   &
+ \kappa_v x_v (1 - p)
\end{eqnarray}

Note that since there is no production of $ vu $ or $ uu $ genomes, we may assume that $ x_{vu} = x_{uu}  = 0 $ at steady-state.  This gives, after normalizing by $ \kappa_{vv} $, the steady-state equations,
\begin{eqnarray}
&  &
0 = -(1 + \phi) x_{vv} + \frac{\gamma \rho}{\kappa_{vv}} x_v^2
\nonumber \\
&  &
0 = -\phi x_v + x_{vv} (1 + p) + \frac{\kappa_v}{\kappa_{vv}} p x_v - \frac{\gamma \rho}{\kappa_{vv}} x_v^2
\nonumber \\
&  &
0 = -\phi x_u + x_{vv} (1 - p) + \frac{\kappa_v}{\kappa_{vv}} (1 - p) x_v
\end{eqnarray}

We will now consider the solution to these equations in the limit where $ \gamma \rho/\kappa_{vv} \rightarrow \infty $.  In this regime, the contact rate between the haploids is so large that $ x_v \rightarrow 0 $ at steady-state.  Adding the first two of the previous set of equations, and setting $ x_v = 0 $, gives,
\begin{equation}
0 = x_{vv} (p - \phi)
\end{equation}  
so that $ \phi = p $ at steady-state.

Clearly, then, the selective mating strategy has a higher mean fitness than the asexual replication strategy, assuming that $ \gamma \rho/\kappa_{vv} \rightarrow \infty $, or equivalently, that $ \kappa_{vv}/(\gamma \rho) \rightarrow 0 $.  

Note that $ \kappa_{vv}/(\gamma \rho) $ measures the ratio of the characteristic fusion time, given by 
$ 1/(\gamma \rho) $, to the characteristic growth time of the cell, given by $ 1/\kappa_{vv} $.  When this ratio is small, that is, when the fraction of time a haploid spends looking for a mate with which to recombine is small compared to the length of the cell cycle, then the steady-state mean fitness of the population is larger than that of the corresponding asexually replicating population.

The quantity $ \kappa_{vv}/(\gamma \rho) $ is referred to as the cost for sex in this model, for when it is small then there is little fitness penalty associated with sexual replication, while when $ \kappa_{vv}/(\gamma \rho) $ is large, then $ \gamma \rho/\kappa_{vv} $ is small, and so haploid fusion is slow, leading to a possibly significant fitness penalty associated with the sexual replication strategy.

What we have established is that, when the cost for sex is negligible, then a selective mating strategy leads to a higher steady-state mean fitness than the asexual strategy.

\subsection{Random mating}

For random mating, the population genetics equations may be readily shown to be,
\begin{eqnarray}
&   &
\frac{d x_{vv}}{dt} = -(\kappa_{vv} + \bar{\kappa}(t)) x_{vv} + \gamma \rho x_v^2
\nonumber \\
&   &
\frac{d x_{vu}}{dt} = -(\kappa_{vu} + \bar{\kappa}(t)) x_{vu} + 2 \gamma \rho x_v x_u
\nonumber \\
&   &
\frac{d x_{uu}}{dt} = -\bar{\kappa}(t) x_{uu} + \gamma \rho x_u^2
\nonumber \\
&   &
\frac{d x_v}{dt} = -\bar{\kappa}(t) x_v + (1 + p) \kappa_{vv} x_{vv} + \frac{1}{2} \kappa_{vu} x_{vu} (1 + p) 
\nonumber \\
&   &
+ \kappa_v x_v p - \gamma \rho x_v (x_v + x_u)
\nonumber \\
&   &
\frac{d x_u}{dt} = -\bar{\kappa}(t) x_u + (1 - p) \kappa_{vv} x_{vv} + \frac{1}{2} \kappa_{vu} x_{vu} (3 - p) 
\nonumber \\
&   &
+ \kappa_v x_v (1 - p) - \gamma \rho x_u (x_v + x_u)
\end{eqnarray}

If we set the left-hand-sides of the above equations to $ 0 $ and normalize by $ \kappa_{vv} $, we obtain that the steady-state equations are,
\begin{eqnarray}
&   &
0 = -(1 + \phi) x_{vv} + (\frac{\gamma \rho}{\kappa_{vv}}) x_v^2
\nonumber \\
&   &
0 = -(\alpha + \phi) x_{vu} + 2 (\frac{\gamma \rho}{\kappa_{vv}}) x_v x_u
\nonumber \\
&   &
0 = -\phi x_{uu} + (\frac{\gamma \rho}{\kappa_{vv}}) x_u^2
\nonumber \\
&   &
0 = -\phi x_v + (1 + p) x_{vv} + \frac{1}{2} \alpha  (1 + p) x_{vu} + \frac{\kappa_v}{\kappa_{vv}} x_v p 
\nonumber \\
&   &
- (\frac{\gamma \rho}{\kappa_{vv}}) x_v (x_v + x_u)
\nonumber \\
&   &
0 = -\phi x_u + (1 - p) x_{vv} + \frac{1}{2} \alpha (3 - p) x_{vu} + \frac{\kappa_v}{\kappa_{vv}} x_v (1 - p) 
\nonumber \\
&   &
- (\frac{\gamma \rho}{\kappa_{vv}}) x_u (x_v + x_u)
\end{eqnarray}

Now, as $ \gamma \rho/\kappa_{vv} \rightarrow \infty $, we obtain that $ x_v, x_u \rightarrow 0 $, and so, adding the last two equations, we obtain,
\begin{equation}
(\frac{\gamma \rho}{\kappa_{vv}}) (x_v + x_u)^2 = 2 \phi
\end{equation}

Defining $ \tilde{x}_v = x_v/(x_v + x_u) $, we then have,
\begin{equation}
(1 + p) (x_{vv} + \frac{1}{2} \alpha x_{vu}) = 2 \phi \tilde{x}_v
\end{equation}

Plugging this expression into the first two equations, we obtain,
\begin{eqnarray}
&   &
\phi (1 + \phi) x_{vv} = (1 + p)^2 \frac{(\phi + x_{vv})^2}{8}
\nonumber \\
&   &
\phi (\alpha + \phi) (\phi - x_{vv}) = \alpha (1 + p) \phi (\phi + x_{vv}) 
\nonumber \\
&   &
- \frac{1}{4} \alpha (1 + p)^2 (\phi + x_{vv})^2
\end{eqnarray}

The first equation may be used to replace $ (\phi + x_{vv})^2 $ in the second equation by a term that is proportional to $ x_{vv} $.  The second equation may then be used for solve for $ x_{vv} $.  The result is,
\begin{equation}
x_{vv} = \phi \frac{\phi - \alpha p}{\phi (1 - 2 \alpha) + \alpha p}
\end{equation}
This expression may then be plugged back in to the first equation, which gives, after some manipulation, that,
\begin{equation}
[\phi + 1] [\phi - \alpha p] [\phi (1 - 2 \alpha) + \alpha p] - \frac{1}{2} (1 + p)^2 (1 - \alpha)^2 \phi^2
= 0
\end{equation}
This expression defines the steady-state mean fitness of the sexual population with random mating, when the cost for sex is negligible.  

When $ \phi = \alpha p $, the left-hand-side evaluates to $ -(1/2) (1 + p)^2 (1 - \alpha)^2 \alpha^2 p^2 \leq 0 $, while when $ \phi = p $, the left-hand-side evaluates to $ (1/2) (1 - p^2) p^2 (1 - \alpha)^2 \geq 0 $, so that $ \phi $ has a solution in $ [\alpha p, p] $, which is the normalized steady-state mean fitness of the population.

When $ p = 0 $ we obtain that $ \phi = 0 $, while when $ p = 1 $ we obtain that $ \phi = 1 $.  When $ \alpha = 0 $ we obtain that $ \phi = \max\{(1/2) (1 + p)^2 - 1, 0\} $, while when $ \alpha = 1 $ we obtain that $ \phi = p $.  

Differentiating both sides of Eq. (22) twice, and substituting $ p = 0 $, gives, after some manipulation,
\begin{equation}
0 = [1 - 2 \alpha - \alpha^2] (\frac{d \phi}{d p})_{p = 0}^2 + 4 \alpha^2 (\frac{d \phi}{d p})_{p = 0} - 2 \alpha^2
\end{equation}
and so,
\begin{equation}
(\frac{d \phi}{d p})_{p = 0} = \alpha (2 \pm \sqrt{2}) \frac{\alpha + 1 \mp \sqrt{2}}{\alpha^2 + 2 \alpha -1}
\end{equation}

Now, $ \alpha^2 + 2 \alpha - 1 = (\alpha + 1 + \sqrt{2}) (\alpha + 1 - \sqrt{2}) $, so that,
\begin{equation}
(\frac{d \phi}{d p})_{p = 0} = \sqrt{2} \alpha \frac{\sqrt{2} \pm 1}{\alpha + 1 \pm \sqrt{2}}
\end{equation}

Since we expect $ \phi $ to be positive for $ p > 0 $, we take the $ + $ solution,
giving,
\begin{equation}
(\frac{d \phi}{d p})_{p = 0} = \alpha \frac{2 + \sqrt{2}}{\alpha + 1 + \sqrt{2}}
\end{equation}

Note that $ (2 + \sqrt{2})/(\alpha + 1 + \sqrt{2}) $ decreases from $ \sqrt{2} > 1 $ to $ 1 $ as $ \alpha $ 
increases from $ 0 $ to $ 1 $.  Therefore, when $ p $ is close to $ 0 $, $ \phi \approx [(2 + \sqrt{2})/(\alpha + 1 + \sqrt{2})] \alpha p > \alpha p $ for $ \alpha < 1 $, so that the mean fitness of the sexual population is greater than that of the asexual population when the replication fidelity is at a low to intermediate value.  

Because $ ([(2 + \sqrt{2})/(\alpha + 1 + \sqrt{2})] - 1) \alpha p $ increases with $ p $, the fitness difference between sexual and asexual replication initially increases as $ p $ increases away from $ 0 $.  If the cost for sex is non-negligible, sexual replication with random mating will only have a selective advantage over asexual replication if this initial fitness difference is sufficiently large.  This implies that sexual replication with random mating will only have a higher mean fitness than asexual replication at intermediate replication fidelities. 

Differentiating both sides of Eq. (22) and substituting $ p = 1 $ gives, after some manipulation,
\begin{equation}
(\frac{d \phi}{d p})_{p = 1} = 2
\end{equation}

As a final note for this subsection, we shall prove that the sexual mean fitness with random mating is greater than the asexual mean fitness when $ r_{12} = r_{14} = 1/2 $, $ r_{13} = 0 $, $ \alpha, p \in (0, 1) $, and there is no cost for sex.

We begin by defining $ \phi_{a} $ to be the normalized mean fitness for the asexual population, and $ \phi_{rs} $ to be the normalized mean fitness for the sexual population with a random mating strategy.

First of all, when $ p = 0 $, then $ \phi_{rs} = \phi_a = 0 $, and when $ p = 1 $, then $ \phi_{rs} = \phi_a = 1 $.  When $ \alpha = 0 $, then $ \phi_a = \phi_{rs} = \max \{2p + r_{12} (1 - p)^2 - 1, 0\} $, while when
$ \alpha = 1 $, then $ \phi_a = \phi_{rs} = p $.  So, in what follows, we will assume that $ \alpha, p \in (0, 1) $.

We have that $ (d \phi_{a}/d p)_{p = 1} = 2 [1 - \alpha (1 - r_{13})]/[1 - \alpha (1 - 2 r_{13})] $, so that, when $ r_{13} > 0 $, $ (d \phi_{a}/d p)_{p = 1} < 2 = (d \phi_{rs}/dp)_{p = 1} $.  This implies that $ \phi_{a} > \phi_{rs} $ for $ p $ sufficiently close, but not equal, to $ 1 $.  Since we also have that $ \phi_{rs} > \phi_{a} $ for $ p $ sufficiently close, but not equal, to $ 0 $, this implies that $ \phi_a $ is initially larger than $ \phi_{rs} $ for $ p $ close to $ 1 $, but then must equal $ \phi_{rs} $ at some $ p \in (0, 1) $ and eventually remains below $ \phi_{rs} $ for $ p $ sufficiently close to $ 0 $.

Now, when $ r_{13} = 0 $, we know that $ \phi_a = \max \{\alpha p, 2 p + r_{12} (1 - p)^2 - 1\} $.  Because $ \phi_{rs} > \alpha p $ for $ \alpha, p \in (0, 1) $, we need only show that $ \phi_{rs} > 2 p + r_{12} (1 - p)^2 - 1 $ for $ r_{12} = 1/2 $.  

Suppose that $ \phi_{rs} = 2 p + (1/2) (1 - p)^2 - 1 = (1/2) (1 + p)^2 - 1 $ for some $ p \in (0, 1) $.  Substituting into Eq. (22) we obtain, after some manipulation, that $ \phi_{rs} = p $.  But then we have that $ (1/2) (1 + p)^2 - 1 = p \Rightarrow p = 1 $, which is a contradiction, since $ p \in (0, 1) $ by assumption.

Therefore, $ \phi_{rs} \neq 2 p + (1/2) (1 - p)^2 - 1 $ for $ \alpha, p \in (0, 1) $, and so $ \phi_{rs} > \phi_a $ whenever $ \alpha, p \in (0, 1) $, $ r_{13} = 0 $, and $ r_{12} = r_{14} = 1/2 $, as we wished to show.

\section{Self-fertilization}

A final replication strategy that we will consider is a form of self-fertilization.  Here, the four haploid cells produced from a given diploid simply fuse with one another.  We consider this replication mechanism independently of whether or not it occurs in {\it S. cerevisiae}.  The reason for this is that we want to examine whether the selective advantage obtained in our sexual replication models could have been obtained without needing to fuse with haploids from a distinct diploid cell.  If it turns out that self-fertilization yields a mean fitness at least as high as the one obtained from our sexual models, then the models we have considered thus far do not provide a clear picture of the selective advantage of sexual replication over asexual replication.

The random mating strategy is equivalent to the asexual replication strategy where $ r_{12} = r_{13} = r_{14} = 1/3 $.  Since we already know that the sexual strategy with random mating will outcompete any asexual strategy at sufficiently low replication fidelities, it follows that the sexual strategy with random mating will outcompete self-fertilization with random mating at sufficiently low replication fidelities.

We now turn to the case of selective mating.  Here, we assume that only two ``v" haploids can fuse with one another.  However, if out of a group of four haploids produced from a given diploid, there is only one ``v" haploid present, we assume that that haploid is lost, and does not participate further in the replication process.

The differential equations governing the time evolution of the population numbers $ n_{vv} $, $ n_{vu} $, and $ n_{uu} $ is then,
\begin{eqnarray}
&   &
\frac{d n_{vv}}{dt} = \kappa_{vv} n_{vv} p^2 + \kappa_{vu} n_{vu} p
\nonumber \\
&   &
\frac{d n_{vu}}{dt} = -\kappa_{vu} n_{vu}
\nonumber \\
&   &
\frac{d n_{uu}}{dt} = 0 
\end{eqnarray}
Note then that when the mutation-selection balance is reached, we have $ n_{vu} = n_{uu} = 0 $, so that, in the long-term, we have $ n = n_{vv} $, and $ d n/dt = \kappa_{vv} p^2 n $, which implies that $ \phi = p^2 < p $ for $ p \in (0, 1) $.  Therefore, self-fertilization with selective mating has a lower mean fitness than sexual replication with selective mating.

\section{Discussion and Conclusions}

The two key results of this paper are that, when the cost for sex is low, the sexual replication model with selective mating has a higher steady-state mean fitness than the asexual replication model, and the sexual replication model with random mating has a higher steady-state mean fitness than the asexual replication model at low to intermediate replication fidelities.  Interestingly, for random mating, the steady-state mean fitness at low to intermediate replication fidelities goes from being equal to the asexual mean fitness when $ \alpha = 1 $, to being a factor of $ \sqrt{2} $ larger than the asexual mean fitness as $ \alpha $ decreases to $ 0 $.

The conclusions of this paper provide a possible explanation for why {\it S. cerevisiae} engages in sexual replication as a stress response:  When conditions are favorable, the replication rate, as measured by $ \kappa_{vv} $, is fairly large, and so the cost for sex, as measured by $ \kappa_{vv}/(\gamma \rho) $, is large as well, so that sex is the disadvantageous strategy.  However, when the yeast population is under stress, $ \kappa_{vv} $ becomes small, and so the cost for sex becomes small as well.  If the mutation rate in yeast is not too low, then sex can become the advantageous strategy.  This explanation is consistent with experimental work on yeast suggesting that sexual replication in yeast provides a mechanism for removing deleterious mutations from the yeast genome \cite{Zeyl:97}.

In a previous series of papers \cite{Tannenbaum:06, TannFon:06}, Tannenbaum and Fontanari considered a different sexual replication strategy than the one being considered here.  Tannenbaum and Fontanari assumed that a given diploid splits directly into two haploids, which enter a haploid pool, fuse, and then the resulting diploid divides asexually.  For both the selective and random mating strategies, it was shown that, when the cost for sex is low and if $ r_{13} = 0 $, sexual replication leads to a higher mean fitness than asexual replication.

Although the replication models considered in \cite{Tannenbaum:06, TannFon:06} led to interesting results, they had two main problems:  First of all, for the random mating strategy, the sexual replication process becomes identical to asexual replication when $ r_{13} = 1 $.  This limited the generality of the conclusions drawn regarding the selective advantage of sexual versus asexual replication in various parameter regimes.

Second, although the models being considered in this paper are highly simplified, they are nevertheless based on a sexual replication pathway that is much closer to the actual sexual replication pathway in {\it S. cerevisiae} (within the constraints of the model).  The replication models considered in \cite{Tannenbaum:06, TannFon:06}, by contrast, were only loosely based on the sexual replication pathway in {\it S. cerevisiae}.  As a result, we argue that the results in \cite{Tannenbaum:06, TannFon:06} could only be used to draw general conclusions about the selective advantage for sexual replication.  On the other hand, we argue that the results in this paper, although based on a highly simplified model, provide a starting point for understanding the selective advantage for sexual versus asexual replication in actual unicellular organisms, e.g. {\it S. cerevisiae}.  

For future research, we would like to develop more realistic models involving multi-gened, multi-chromosomed genomes.  Furthermore, this paper explored the selective advantage for sex in static environments.  For future research, we would like to develop models describing competition between sexual and asexual replication in dynamic environments, to test the theories for sex that are based on the idea that sex increases adaptability.  

In this vein, it should be noted that there has been experimental work on yeast suggesting that sex is advantageous because it allows yeast to adapt more quickly to changing environments \cite{Goddard:05}.  As we have mentioned previously, this view is not necessarily contradictory to the results of \cite{Zeyl:97}, which argue that sex is a way of removing deleterious mutations.  In our opinion, at intermediate mutation rates, and when the cost for sex is sufficiently low, there is an advantage for sex that holds in static environments and in an infinite population.  This advantage, however, can be significantly enhanced in dynamic environments, and for intermediate population sizes \cite{Otto:06}.

\begin{acknowledgments}

This research was supported by the United States - Israel Binational Science Foundation (Start-Up Grant), and by the Israel Science Foundation (Alon Fellowship).  The author would also like to thank J.F. Fontanari for suggesting this research problem.

\end{acknowledgments}

\end{document}